# Full Paper

# Thermal conductivity tensor of $\beta$-HMX as a function of pressure and temperature from equilibrium molecular dynamics simulations

Andrey Pereverzev*[a] and Tommy Sewell‡[a]

**Abstract:** We apply the Green–Kubo (G-K) approach to obtain the thermal conductivity tensor of $\beta$-1,3,5,7-tetranitro-1,3,5,7-tetrazoctane ($\beta$-HMX) as a function of pressure and temperature from equilibrium molecular dynamics (MD) simulations. Direct application of the G-K formula exhibits slow convergence of the integrated thermal conductivity values even for long (120 ns) accumulated simulation times. To partially mitigate this slow convergence, we apply a recently implemented numerical procedure [Int. J. Heat Mass Trans. **188**, 122647 (2022)] that involves physically justified filtering of the MD-calculated G-K heat current and fitting the integrated time-dependent thermal conductivity to a physically motivated double-exponential function. The thermal conductivity tensor was determined for pressures 1 atm $\leq P \leq$ 30 GPa and temperatures 300 K $\leq T \leq$ 900 K. The thermal conductivity $\kappa^{\alpha\beta}(T,P)$ increases with increasing pressure, by approximately an order of magnitude over the interval considered, and decreases with increasing temperature. The MD predictions are compared to experimental and other theoretically determined values for the thermal conductivity of $\beta$-HMX. A simple, semi-empirical fitting form is proposed that captures the behavior of $\kappa^{\alpha\alpha}(T,P)$ over the pressure and temperature intervals studied.

# Full Paper

## I. INTRODUCTION

Understanding the response of a heterogeneous explosive (HE) material subjected to a thermo-mechanical insult is a longstanding challenge to the energetic materials community [ 1 , 2 ]. Substantial effort has been devoted to predicting the response of HE to shock stimuli, with increased attention in recent years on the development of accurate grain-resolved continuum models and simulations that account explicitly for mesoscale physics and material microstructure. See, for example, Ref. [3].

Passage of a shock wave through the microstructure of a heterogeneous explosive (HE) leads to localized regions of high temperature, stress, and strain rate behind the shock. These regions are known colloquially as hotspots and are thought to be critical to triggering chemistry in the charge. For a given hotspot, the competition between local chemical energy release and energy dissipation due to thermal conduction will largely determine whether the hotspot will quench or grow. Hotspots that grow will eventually interact with others, possibly culminating in violent explosion or detonation [1,2].

The need to understand and characterize energy localization and hotspot interactions theoretically across scales is particularly acute in scenarios involving impacts or weak shocks, for which an undesired deflagration-to-detonation transition can lead to unfortunate outcomes. Whereas for strong shocks pressure overwhelms strength such that the deformation and flow are hydrodynamic-like, under weak loading the post-shock material response is dominated by the solid equation of state, transport properties, and crystal and interfacial properties and dynamics. Thus, physics-based prediction of near-threshold impact events requires accurate knowledge of these properties—essentially all of which are anisotropic for explosive polymorphs—across the wide intervals of temperature, pressure, and in some cases strain rate that are relevant to detonation initiation.

The thermal conductivity of the constituent materials is one of the key inputs needed for such modelling efforts. Experimentally determined thermal conductivities for explosives are subject to large uncertainties associated with measurement techniques and sample preparation (see, for example, Table I of Ref. [4]). With rare exceptions [5-8], experimental data for elevated pressures and temperatures are not available even for pressed powders. Determinations of the thermal conductivity tensor for single oriented high-explosive crystals under any conditions are especially rare [7]. Fortunately, molecular dynamics (MD) simulations are well suited for providing thermal conductivity values and other fundamental information needed by the mesoscale models [3,9-15].

[a]    A. Pereverzev, T. Sewell
       *Department of Chemistry and Materials Science*
       *& Engineering Institute*
       *University of Missouri*
       *Columbia, MO 65211, USA*
       *E-mails: pereverzeva@missouri,edu,*
       *sewellt@missouri.edu*

The MD simulation techniques used to obtain the thermal conductivity tensor of crystalline solids fall into two main categories: reverse non-equilibrium molecular dynamics (rNEMD) [16] and the equilibrium Green-Kubo (G-K) [17] approach. Each category has certain advantages and disadvantages. In the rNEMD approach, a heat current is imposed on the system in a given direction, resulting in a temperature gradient. The current and the gradient are then used to calculate the directional thermal conductivity using Fourier's law. To obtain the full thermal conductivity tensor at a given $(T, P)$ state, rNEMD requires setting up as many simulations as there are independent nonzero components of the thermal conductivity tensor [18]. By contrast, G-K yields the full tensor from a single equilibrium simulation. Generally, rNEMD simulations require adjustment of empirical simulation parameters to achieve a suitable thermal gradient. These include simulation-cell length and cross-sectional area, hot and cold region widths, velocity swapping frequency, and how the velocity swapping is performed (i.e., based on individual atoms or molecular centers of mass). In the case of G-K performed in the NVE ensemble there are no adjustable parameters other than the system size, which can be assessed using Mathiessen's rule and used to extrapolate the predicted conductivity to infinite system size. However, the G-K method often requires long simulation times to obtain well-converged results and finite-size effects are often significant. The rNEMD approach has been applied to predict the thermal conductivities of several different energetic materials [4,7,18-24]. The G-K approach was used to determine the thermal conductivity of $\alpha$-RDX [19] and $\beta$-HMX [25] at standard ambient conditions.

Most MD studies of thermal conductivity of $\beta$-HMX considered either the liquid state [26] or did not specifically include temperature and pressure effects [4,20,25]. Recently, Perriot and Cawkwell [18] studied the pressure and temperature dependence of the $\beta$-HMX thermal conductivity tensor for 200 K ≤ $T$ ≤ 500 K and 0 GPa ≤ $P$ ≤ 5 GPa. Experimentally, a few studies

# Full Paper

were performed on β-HMX polycrystals/powders (sometimes porous) that did not resolve the effect of crystal orientation [6,8,27]. Others worked with β-HMX-based plastic-bonded explosives (PBXs) [5,8]. Hanson-Parr and Parr [6] studied the effect of temperature on the thermal conductivity of pressed powders for 293.15 K ≤ T ≤ 433.15 K. Lawless et al. [8] performed experimental determination of thermal conductivity of pressed powder β-HMX and some β-HMX-based PBXs at T = 300 K and 441 K. Cornell and Johnson [5] measured thermal conductivity of several β-HMX-based PBXs for 228 K ≤ T ≤ 350 K. Each of these studies will be discussed further in Sec. III.

In a recent study [25], we used the G-K approach to predict the thermal conductivity tensor of β-HMX. The main focus of that study was on methods development, and only the (300 K, 1 atm) state was considered. Here, we apply the G-K approach as described in Ref. [25] to obtain the thermal conductivity tensor of β-HMX for pressures 1 atm ≤ P ≤ 30 GPa and temperatures 300 K ≤ T ≤ 900 K. These intervals extend significantly the regions of (T, P) space considered in the other studies [5,6,8,18].

## II. SIMULATION DETAILS

All MD simulations were performed using the LAMMPS [28] package. We employed the nonreactive, fully flexible molecular potential-energy surface (force field) for HMX proposed by Smith and Bharadwaj [29] and further developed by Bedrov et al. [30] and others [3,31]. This force field is well-validated and has been used in numerous previous studies of HMX [4,18, 25, 26, 30,32-35]. We used the version of the force field first described in Ref. [3]. Sample LAMMPS input decks including all force-field parameters, details of how the forces were evaluated, and a Cartesian crystal supercell description can be found in the supplementary material of Ref. [25].

Simulations were performed for three-dimensionally (3D) periodic supercells of β-HMX comprising $8a \times 8b \times 8c$ unit cells, where $a$, $b$, and $c$ are the unit-cell lattice vectors in the P2₁/n space group setting. The mapping between the crystal and Cartesian lab frame is $a \parallel \hat{x}$, $b \parallel \hat{y}$, and $c$ in the $+\hat{z}$ half space. This system size was chosen based on our previous study [25] of size-dependence of thermal conductivity of β-HMX at T = 300 K and P = 1 atm, as a deliberate compromise between computational efficiency and the need to calculate conductivity for the bulk (infinite-size) crystal. Estimates in Ref. [25] based on Mathiessen's rule suggest that the thermal conductivity for an $8a \times 8b \times 8c$ supercell is approximately 12% smaller than that for the bulk crystal. We note that the deviations of the thermal conductivity tensor at high T and P for the $8a \times 8b \times 8c$ supercell relative to that for the bulk crystal may differ from that at standard ambient conditions, as phonon mean free paths depend on both T and P and these can affect the system-size dependence of thermal conductivity. In this work we *assume* that the results for the $8a \times 8b \times 8c$ supercell are sufficiently close to the bulk crystal limit not only for (300 K, 1 atm) but for higher (T, P) states as well. *Vide infra*, this assumption appears to be reasonable, as the predicted conductivity values span approximately an order of magnitude on the (T, P) interval studied.

As part of our methods-development study [25] we obtained the β-HMX thermal conductivity tensor for P = 1 atm and T = 300 K. Here, we extend the interval of temperatures and pressures as follows: For each of the four higher pressures studied——P = 5, 10, 20, and 30 GPa—— we consider T = 300, 500, and 900 K. For P = 1 atm, we consider T = 300 (from Ref. [25]), 500, and 700 K. These overall intervals coincide with the ones used in Ref. [35], where we reported the T - and P -dependence of the β-HMX elastic tensor. The upper temperature limit for states on the P = 1 atm isobar was set to 700 K (instead of 900 K as for the elevated pressures) because the crystal melted during MD simulations performed at (900 K, 1 atm), whereas the (metastable, superheated) crystal remained intact, on the time scale of the simulations, at (700 K, 1 atm).

For each (T, P) pair, the supercell lattice parameters were determined as arithmetic averages over the final 500 ps of 600 ps long isobaric-isothermal (NPT) trajectories, sampled every 100 fs, in which all six lattice parameters were allowed to vary independently (LAMMPS *tri* keyword). The target stress state was a multiple of the unit tensor, the integration time step was 0.2 fs, and the barostat and thermostat coupling parameters were set to 200 fs and 20 fs, respectively. The resulting cell parameters, which agree well with those tabulated in our recent study of β-HMX (T, P)-dependent elasticity [35], were used for 500 ps long isochoric-isothermal (NVT) trajectories at corresponding temperatures, with velocity re-selections at 100 ps intervals, to prepare the supercells for production trajectories. The NVT time step and thermostat coupling parameter were 0.2 fs and 20 fs, respectively.

The thermal conductivity tensor $\kappa^{\alpha\beta}$ in the G-K approach is calculated from the following relationship [17],

$$\kappa^{\alpha\beta} = \frac{1}{k_B T^2 V} \int_0^\infty dt \langle J^\alpha(0) J^\beta(t) \rangle, \qquad (1)$$

# Full Paper

where $J^\alpha$ is the $\alpha^{th}$ Cartesian component of the heat current, $V$ is the system volume, and the brackets indicate averaging over an equilibrium ensemble. Because $\beta$-HMX is a monoclinic crystal and given the chosen crystal-frame/lab-frame mapping, only the following components of the thermal conductivity tensor are nonzero: $\kappa^{xx}$, $\kappa^{yy}$, $\kappa^{zz}$, $\kappa^{xz}$, and $\kappa^{zx}$. For the $\kappa^{xz}$ and $\kappa^{zx}$ tensor components, which are formally equivalent, we report the numerical average of the two components and label it with the superscript $xz$.

To obtain the thermal conductivity tensor using Eq. (1), the heat current was calculated using ensembles of 30 independent isochoric-isoenergetic (NVE) trajectories for each $(T, P)$ state. Each trajectory was 4.0 ns long and the heat-current data were recorded at 1.0 fs intervals. A time step of 0.1 fs was used.

The heat current for a system consisting of $N$ atoms as calculated by LAMMPS is defined by the following expression:

$$\mathbf{J} = \sum_{i=1}^{N} \epsilon_i \mathbf{v}_i - \sum_{i=1}^{N} \mathbf{S}_i \cdot \mathbf{v}_i, \qquad (2)$$

where $\epsilon_i$ is the energy of atom $i$ and $\mathbf{S}_i$ is the per-atom stress tensor of that atom [28]. The tensor $\mathbf{S}_i$ in Eq. (2) multiplies atomic velocity $\mathbf{v}_i$ as a $3 \times 3$ matrix multiplies a vector, to yield a vector.

To calculate the heat current using Eq. (2), LAMMPS first computes the per-atom stress tensor $\mathbf{S}_i$ [28]. The version of LAMMPS (October 29, 2020) used for the present simulations can calculate two different types of per-atom stress tensor, specified by using the *centroid/stress/atom* and *stress/atom* keywords. Recently, it was shown that the *centroid/stress/atom* keyword must be used to correctly calculate the heat flux for systems with three- and four-body interactions [36]. However, the version of LAMMPS used here does not support the *centroid/stress/atom* keyword for potentials with long-range Coulombic interactions, which are present in the force field. To overcome this limitation, we used the following hybrid approach to calculate the heat current [25]: The *stress/atom* keyword was used for the part of the current arising from all two-body interactions (i.e., covalent bonds and non-bonded terms including Coulombic) and the *centroid/stress/atom* keyword was used for the part of the current due to three- and four-body interactions (i.e., for bond angles, dihedrals, and improper dihedrals). The two parts were then added to obtain the total *hybrid* heat current.

This hybrid heat current was filtered to reduce the oscillations of the heat-current correlation functions (HCCFs). The filtering, first proposed in Ref. [37] by Marcolongo et al. and applied by us to $\beta$-HMX in Ref. [25], involves subtracting the linear function of atomic velocities given by $\sum_{i=1}^{N}(\frac{5kT}{2} + \langle u_i \rangle) \mathbf{v}_i - \sum_{i=1}^{N} \langle \mathbf{S}_i \rangle \cdot \mathbf{v}_i$ from the hybrid heat current given by Eq. (2). Here, $\langle u_i \rangle$ and $\langle \mathbf{S}_i \rangle$ are, respectively, the ensemble-averaged per-atom potential energy and stress. It was shown in Ref. [38] that subtracting such a linear function of the velocities from the heat current (2) does not affect thermal conductivity calculated using the G-K approach but generally reduces the oscillations in the HCCF. The filtering scheme is discussed in greater detail in Ref. [25]. That reference considered two filtering schemes. Here, we used the second of the two.

The HCCFs, defined as

$$C^{\alpha\beta}(t) = \langle J^\alpha(0) J^\beta(t) \rangle, \qquad (3)$$

were calculated using the filtered hybrid heat current components $J^\alpha$. These correlation functions were then used to obtain time-dependent G-K thermal conductivities given by

$$\kappa^{\alpha\beta}(t) = \frac{1}{k_B T^2 V} \int_0^t d\tau \, \langle J^\alpha(0) J^\beta(\tau) \rangle. \qquad (4)$$

An example of typical behavior of the time-dependent thermal conductivities is shown in Fig. 1. One can see that, even using a total simulation time of 120 ns combined with the hybrid heat current and heat-current filtering, the time-dependent thermal conductivity components are not fully converged to a definite, constant value. To obtain definite values of the thermal conductivity for each diagonal tensor component, we used the fitting function

$$F(t) = B_1 + B_2 - B_1 \exp(-\gamma_1 t) - B_2 \exp(-\gamma_2 t), \quad (5)$$

where $B_1$, $B_2$, $\gamma_1$, and $\gamma_2$ are fitting parameters. The asymptotic value of a given $\kappa^{\alpha\alpha}$ is then given by $B_1 + B_2$. Physically, the third and fourth terms in Eq. (5) account for the contributions to $\kappa^{\alpha\beta}(t)$ from the acoustic phonons with long and short wavelengths, respectively [39]. More detailed discussion of the fitting procedure is given in Ref. [25].

The function (5) is fitted to the time-dependent thermal conductivity data for each diagonal component in the interval where the $\kappa^{\alpha\alpha}(t)$ exhibit monotonic increase; that is, where the time-dependent data are consistent with Eq. (5). Our data analysis revealed that the interval of monotonic increase of $\kappa^{\alpha\alpha}(t)$ depends on pressure but is essentially independent of temperature. Consequently, the following empirically determined pressure-dependent fitting intervals were used: from 2 to 30 ps for

# Full Paper

states on the $P = 1$ atm isobar, 2-40 ps for $P = 5$ GPa, 2-50 ps for $P = 10$ and 20 GPa, and 2-60 ps for $P = 30$ GPa. The off-diagonal component $\kappa^{xz}(t)$ is much smaller in absolute value compared to the diagonal components and does not exhibit a clear exponential-type behavior. Therefore, for $\kappa^{xz}$ we report the values of the time integrals taken at 10 ps, for which $\kappa^{xz}(t)$ becomes approximately constant while the estimated uncertainty remains much smaller than the absolute value.

## III. RESULTS AND DISCUSSION

The thermal conductivity tensor components as functions of pressure and temperature are listed in Table I. Also listed is the average thermal conductivity, $\bar{\kappa} = (\kappa^{xx} + \kappa^{yy} + \kappa^{zz})/3$ . The pressure and temperature dependencies of the diagonal components of the thermal conductivity tensor and $\bar{\kappa}$ are depicted graphically in Figs. 2 and 3.

Table I and Fig. 2 show that the thermal conductivity tensor is strongly dependent on pressure. Between 1 atm and 30 GPa, $\kappa^{xx}$ and $\kappa^{zz}$ increase approximately ninefold, $\kappa^{yy}$ increases approximately sixfold, and $\bar{\kappa}$ grows by a factor of eight, for both $T = 300$ K and $T = 500$ K. It can be seen from Fig. 2 that the pressure dependence for the diagonal tensor components is, in general, nonlinear. Increase of the thermal conductivity with pressure was also observed in MD studies of $\beta$-HMX [18], TATB [22], $\alpha$-RDX [7], PETN [23], and γ and ε-HNIW [24]. Note however that excepting the PETN study in Ref. [23], which extended to 8 GPa, all of these were limited to $P \leq 5$ GPa for which approximately linear pressure dependencies of the thermal conductivity were observed. On the sixfold-wider pressure interval studied here, the non-linear $P$-dependence is readily apparent.

The temperature dependence of the diagonal tensor components (see Fig. 3) exhibits an approximately linear relationship between each component and $1/T$. Note that the slopes of the linear fits in Fig. 3 increase with increasing pressure. For $P = 1$ atm the average thermal conductivity decreases by approximately a factor of two in the temperature interval between 300 and 700 K. For all higher pressures the average thermal conductivity decreases by approximately the same factor, but across the wider interval 300 $\leq T \leq$ 900 K. An approximately linear dependence of the thermal conductivity on $1/T$ was reported for MD simulations of $\beta$-HMX [18], TATB [22], $\alpha$-RDX [7], PETN [23], and γ and ε-HNIW [24]. In these references it was argued that such behavior agrees with the classical limit of theoretical models [ 40 - 42 ]. However, those theoretical models predict a temperature dependence of the form $\kappa^{\alpha\beta} = m/T$ (i.e., $\lim\limits_{T\to\infty} \kappa^{\alpha\beta} = 0$); whereas the dependence observed in Refs. [7,18,22-24] as well as here is of the form $\kappa^{\alpha\beta} = a + m'/T$. Additional theoretical analysis, which will not be pursued here, is necessary to explain this discrepancy.

Reference [25] contains comparisons among measured and predicted values for the thermal conductivity tensor at $P = 1$ atm and $T = 300$ K. These comparisons reveal significant differences in the thermal conductivity values both among different experimental data and among different MD approaches. In the following, we compare our results for higher pressures and temperatures to literature values which, unfortunately, are very scarce.

Hanson-Parr and Parr [6] measured the $P = 1$ atm thermal diffusivity and heat capacity of pressed-powder HMX for temperatures 293.15 K $\leq T \leq$ 433.15 K and obtained the thermal conductivity as a linear function of temperature. Extrapolating their linear function yields $\kappa = 0.39$ W m$^{-1}$ K$^{-1}$ at $T = 500$ K and $\kappa = 0.29$ W m$^{-1}$ K$^{-1}$ at $T = 700$ K. These values are approximately 88% of what we obtain for $\bar{\kappa}$ at the respective temperatures. That our values are larger is consistent with theoretical expectations, as our systems do not contain any free surfaces (due to the use of periodic boundary conditions) and do not include internal defect structures such as stacking faults or grain boundaries, all of which will certainly exist in pressed powders and will reduce the conductivity due to phonon scattering effects. However, given the extrapolation of Hanson-Parr and Parr's linear fit to data and uncertainty as to force field accuracy in the MD, it is difficult to offer a more incisive explanation for the (rather modest) discrepancy.

Lawless et al. [8] measured the thermal conductivity of $\beta$-HMX powder, pressed to 90% of the theoretical maximum density, at $T = 300$ K and 441 K. Thus, their results cannot be directly compared to ours. Somewhat surprisingly, they did not observe any difference in the conductivity at these two temperatures and therefore reported the single value, $\kappa = 0.35$ W m$^{-1}$ K$^{-1}$, for both. Those authors also did not observe any effects of temperature on the thermal conductivity of several different $\beta$-HMX-based PBXs, based on measurements performed at those same two temperatures.

In another experimental study, Cornell and Johnson [5] measured the thermal conductivity of several $\beta$-HMX-based PBXs for 228 K $\leq T \leq$ 350 K. While their data are not directly comparable to ours, they did report a

# Full Paper

pronounced decrease of the thermal conductivity with increasing temperature.

Perriot and Cawkwell [18] calculated the thermal conductivity tensor of $\beta$-HMX, for 200 K $\leq T \leq$ 500 K and 0 GPa $\leq P \leq$ 5 GPa, using rNEMD simulations in conjunction with the Smith-Bharadwaj force field. These pressure and temperature intervals partially overlap with the ones considered here. The results of Perriot and Cawkwell for $T$ = 300 and 500 K and $P$ = 0 and 5 GPa are included in Table I. Our values for the diagonal components of the thermal conductivity tensor are consistently higher than those obtained by Perriot and Cawkwell. Whereas our values for $\kappa^{xz}$ are always negative, Perriot and Cawkwell report both positive and negative values for that component. The reasons for these differences are not clear, as the force fields used in Ref. [18] and here are very nearly the same. The known distinctions are: (1) different N–O and C–H covalent bond-stretching force constants; (2) incorporation of a steep, very-short-range, purely repulsive non-bonded interatomic pair potential in the present study; and (3) omission in the present study of a small potential-energy term $U(\phi)$ for the O-N-N-C dihedral angle $\varphi$, which depends on $\varphi$ as cos($8\varphi$). Perriot and Cawkwell discuss possible implications of the differing force constants in some detail. Regarding the steep repulsive core, given the very short distance interval for which it is practically non-zero, we are confident that it has little if any effect on the present predictions. (The Supporting Information for a recent study due to Kroonblawd and Springer [43] includes a detailed exploration of the effects of the same short-range repulsive potential on various properties computed for $\alpha$-RDX using essentially the same force field as employed here. Their results reinforce our confidence that the repulsive potential is not the cause of the discrepancy between the present predictions and those of Perriot and Cawkwell.) The same is true for the third distinction: the maximum difference in internal-coordinate forces, $-\partial U(\phi_{ONNC})/\partial \phi_{ONNC}$, for the two slightly differing forms of the O-N-N-C dihedral angle is less than 1% everywhere on the angular domain. Thus, the explanation of why ostensibly equivalent approaches, simulated using such similar force fields, yields such different results for the thermal conductivity is an outstanding question that deserves further attention. If the force constants do not explain the difference, it would appear that it arises due to differences between the rNEMD and G-K methods.

In addition to the differences in the thermal conductivity values discussed immediately above, the $T$ and $P$ dependence of the thermal conductivity tensor reported in Ref.

[18] is not fully consistent with our results. Perriot and Cawkwell [18] proposed the following functional form for each nonzero tensor component $\kappa^{\alpha\beta}(T, P)$:

$$\kappa^{\alpha\beta}(T, P) = \kappa_0^{\alpha\beta} + \frac{a_1^{\alpha\beta}}{T} + a_2^{\alpha\beta} P, \qquad (6)$$

where $\kappa_0^{\alpha\beta}$, $a_1^{\alpha\beta}$, and $a_2^{\alpha\beta}$ are fitting parameters that are independent of $T$ and $P$. As discussed above, our results exhibit a $1/T$ temperature dependence for the diagonal components of the conductivity tensor. However, as can be seen in Fig. 3, the slopes of those linear fits increase with pressure. This is noticeable even when pressure increases from 1 atm to 5 GPa, that is, in the pressure interval studied in Ref. [18]. Such temperature dependence of our results implies that, if we accept the validity of Eq. (6), the fitting parameter $a_1^{\alpha\beta}$ has to be pressure dependent. Also note that the pressure dependence of our results (see Fig. 2) is clearly nonlinear and therefore is not accurately captured by the simple linear dependence on $P$ given by Eq. (6). The fact that Eq. (6) provides a good description of the thermal conductivity values obtained by Perriot and Cawkwell is likely due to the relatively narrow interval of pressures considered by them, which is only 1/6 of that studied here. In his MD study of thermal conductivity of PETN [23], which considered pressures 0 GPa $\leq P \leq$ 8 GPa, Perriot proposed to add a $P^2$-dependent term to Eq. (6) to better fit his numerical results.

Equation (6) can be treated as a two-variable Taylor-series expansion of the thermal conductivity tensor in powers of $1/T$ and $P$, truncated at the first-order (linear) terms. To obtain an expression that better fits the data for broader intervals in the $(T, P)$ space, we extended the Taylor series to the second order in $1/T$ and $P$:

$$\kappa^{\alpha\beta}(T, P) = \kappa_0^{\alpha\beta} + \frac{a_1^{\alpha\beta}}{T} + a_2^{\alpha\beta} P + \frac{a_3^{\alpha\beta}}{T^2}$$
$$+ a_4^{\alpha\beta} P^2 + \frac{a_5^{\alpha\beta} P}{T}. \qquad (7)$$

Equation (7) was fit to the full set of conductivity data by using the default Levenberg-Marquadt algorithm in Mathematica [44]. In the fits, each data point was weighted by a factor $1/\sigma^2$, where $\sigma$ is the uncertainty associated with the datum. The fits yield reasonably good agreement with the data, with an average relative error below 4%. As an example, Fig. 4 shows the MD results for $\bar{\kappa}(T, P)$ along with the fitted function [Eq. (7)]. No fits were done for $\kappa^{xz}(T, P)$ tensoral component because, in contrast to the diagonal components, this component does not

# Full Paper

itself represent directional thermal conductivity. The fitted parameters are listed in Table II.

## IV. CONCLUSIONS

We used G-K MD simulations and a variant of the Smith–Bharadwaj nonreactive force field to determine the full thermal conductivity tensor of $\beta$-HMX, $\kappa^{\alpha\beta} = \kappa^{\alpha\beta}(T, P)$, for 300 K $\leq T \leq$ 900 K and 1 atm $\leq P \leq$ 30 GPa. The thermal conductivity is predicted to increase by nearly an order of magnitude on that pressure interval, monotonically but with a nonlinear pressure dependence. At a given pressure, $\kappa^{\alpha\alpha}$ is predicted to decrease with temperature as $1/T$, which is partially but not fully consistent with theoretical expectations. At a given temperature, the slope of $\kappa^{\alpha\beta}$ vs. $P$ increases with increasing pressure. Fits of a second-order Taylor expansion of $\kappa^{\alpha\beta}(T, P)$ in the variables $P$ and $1/T$ [Eq. (7)] yield an analytic, semi-empirical model that describes the calculated $T$- and $P$-dependence of $\kappa^{\alpha\beta}$ with acceptable accuracy.

*Acknowledgements*

This research was funded by Air Force Office of Scientific Research, Grant Nos. FA9550-19-1-0318 and FA9550-22-1-0212; and by an AFOSR DURIP equipment award, Grant No. FA9550-20-1-0205. The authors are grateful to Romain Perriot and Marc Cawkwell for sharing their numerical data.

# Full Paper

# Full Paper

# Full Paper

TABLE I: Predicted components of the thermal conductivity tensor $\kappa^{\alpha\beta}$ and isotropic thermal conductivity $\bar{\kappa}$ of $\beta$-HMX as functions of pressure and temperature. Results of Ref. 18 are included for comparison. Units are W m$^{-1}$ K$^{-1}$. Here and elsewhere in this paper, uncertainties reflect one standard deviation about the respective mean values.

| $P$ (GPa) | $T$ (K) | $\kappa^{xx}$ | $\kappa^{yy}$ | $\kappa^{zz}$ | $\kappa^{xz}$ | $\bar{\kappa}$ |
|---|---|---|---|---|---|---|
| $10^{-4}$ | 300 | 0.68 ± 0.02 | 0.69 ± 0.02 | 0.52 ± 0.02 | -0.014 ± 0.06 | 0.63 ± 0.01 |
| | 300[a] | 0.541 ± 0.015 | 0.562 ± 0.008 | 0.392 ± 0.013 | 0.010 ± 0.028 | 0.498 ± 0.007 |
| | 500 | 0.47 ± 0.01 | 0.49 ± 0.02 | 0.38 ± 0.01 | −0.032 ± 0.004 | 0.44 ± 0.01 |
| | 500[a] | 0.436 ± 0.010 | 0.433 ± 0.013 | 0.321 ± 0.015 | −0.024 ± 0.015 | 0.397 ± 0.007 |
| | 700 | 0.320 ± 0.009 | 0.37 ± 0.01 | 0.301 ± 0.008 | −0.038 ± 0.004 | 0.33 ± 0.01 |
| 5 | 300 | 2.12 ± 0.07 | 1.45 ± 0.04 | 1.54 ± 0.04 | −0.20 ± 0.02 | 1.71 ± 0.05 |
| | 300[a] | 1.153 ± 0.022 | 0.890 ± 0.006 | 0.932 ± 0.031 | 0.165 ± 0.027 | 0.992 ± 0.013 |
| | 500 | 1.48 ± 0.05 | 1.09 ± 0.04 | 1.08 ± 0.03 | −0.14 ± 0.02 | 1.22 ± 0.04 |
| | 500[a] | 1.037 ± 0.022 | 0.81 ± 0.04 | 0.815 ± 0.033 | −0.04 ± 0.04 | 0.887 ± 0.018 |
| | 900 | 0.95 ± 0.03 | 0.78 ± 0.03 | 0.92 ± 0.03 | −0.16 ± 0.01 | 0.89 ± 0.03 |
| 10 | 300 | 3.0 ± 0.1 | 2.26 ± 0.08 | 2.3 ± 0.1 | −0.32 ± 0.02 | 2.5 ± 0.1 |
| | 500 | 2.11 ± 0.08 | 1.58 ± 0.06 | 1.69 ± 0.07 | −0.27 ± 0.02 | 1.79 ± 0.07 |
| | 900 | 1.45 ± 0.05 | 1.19 ± 0.05 | 1.24 ± 0.04 | −0.27 ± 0.01 | 1.29 ± 0.05 |
| 20 | 300 | 4.6 ± 0.2 | 3.0 ± 0.1 | 3.5 ± 0.2 | −0.46 ± 0.04 | 3.7 ± 0.1 |
| | 500 | 3.3 ± 0.1 | 2.2 ± 0.1 | 2.77 ± 0.09 | −0.39 ± 0.03 | 2.7 ± 0.1 |
| | 900 | 2.18 ± 0.07 | 1.71 ± 0.09 | 1.84 ± 0.08 | −0.32 ± 0.03 | 1.91 ± 0.08 |
| 30 | 300 | 6.1 ± 0.3 | 4.0 ± 0.2 | 4.9 ± 0.2 | −0.61 ± 0.05 | 5.0 ± 0.2 |
| | 500 | 4.0 ± 0.2 | 3.0 ± 0.1 | 3.3 ± 0.2 | −0.48 ± 0.04 | 3.5 ± 0.2 |
| | 900 | 2.9 ± 0.1 | 2.1 ± 0.1 | 2.30 ± 0.09 | −0.38 ± 0.03 | 2.4 ± 0.1 |

[a] Ref. 18.

# Full Paper

TABLE II: Parameters obtained from fitting Eq. (7) to the MD thermal conductivity predictions.

| | $\kappa^{xx}$ | $\kappa^{yy}$ | $\kappa^{zz}$ | $\bar{\kappa}$ |
|---|---|---|---|---|
| $\kappa_0{}^a$ | -0.201 | -0.02 | 0.135 | -0.006 |
| $a_1$ | 438 | 316 | 121 | 270 |
| $a_2$ | 0.1003 | 0.0788 | 0.0830 | 0.0912 |
| $a_3$ | -50761 | -30208 | -729 | -23492 |
| $a_4$ | -0.00234 | -0.00149 | -0.00190 | -0.00204 |
| $a_5$ | 46.1 | 22.6 | 35.5 | 33.6 |

[a] Units for $\kappa_0$, $a_1$, $a_2$, $a_3$, $a_4$, and $a_5$ are, respectively, W m$^{-1}$ K$^{-1}$, W m$^{-1}$, W m$^{-1}$ K$^{-1}$ GPa$^{-1}$, W m$^{-1}$ K, W m$^{-1}$ K$^{-1}$ GPa$^{-2}$, and W m$^{-1}$ GPa$^{-1}$.

# Full Paper

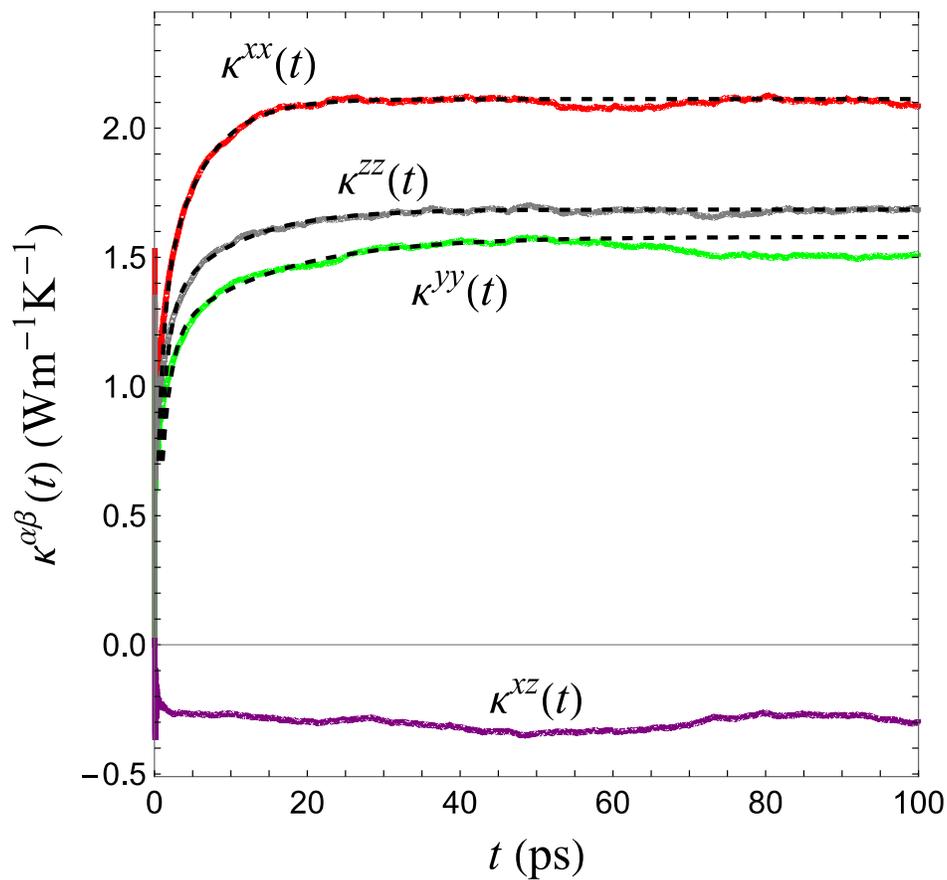

FIG. 1. Time-dependent thermal conductivity tensor components $\kappa^{xx}$, $\kappa^{yy}$, $\kappa^{zz}$, and $\kappa^{xz}$ of $\beta$-HMX for the case $P = 10$ GPa and $T = 500$ K. The dashed black curves for the diagonal components are fits of the double-exponential function [Eq. (5)] to the MD data.

# **Full Paper**

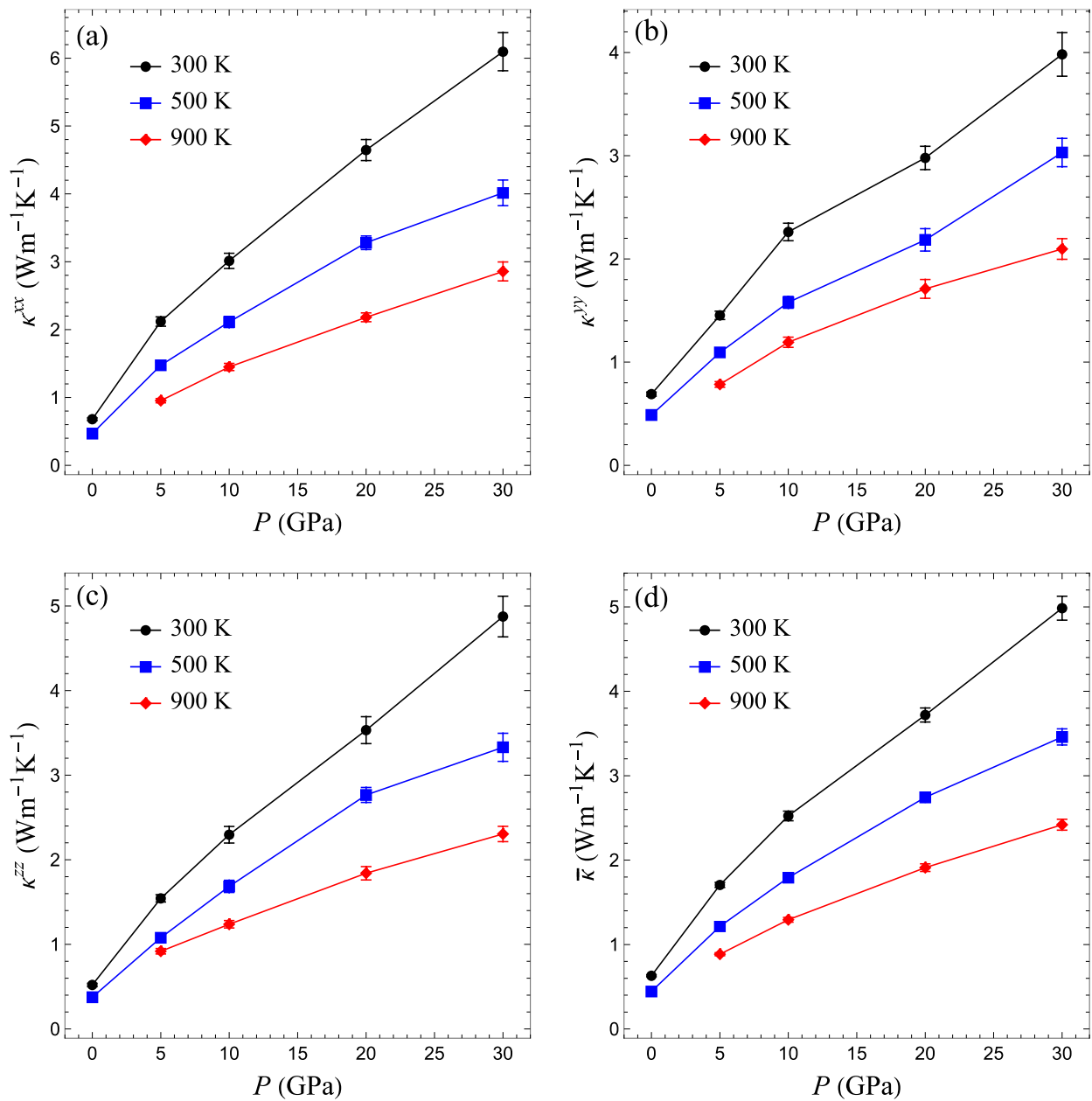

FIG. 2. (a) Thermal conductivity tensor component $\kappa^{xx}$ of $\beta$-HMX as a function of pressure on the $T = 300$, 500, and 900 K isotherms. Solid curves serve as guides to the eye. (b) Same as (a) but for $\kappa^{yy}$. (c) Same as (a) but for $\kappa^{zz}$. (d) Same as (a) but for $\overline{\kappa}$.

# **Full Paper**

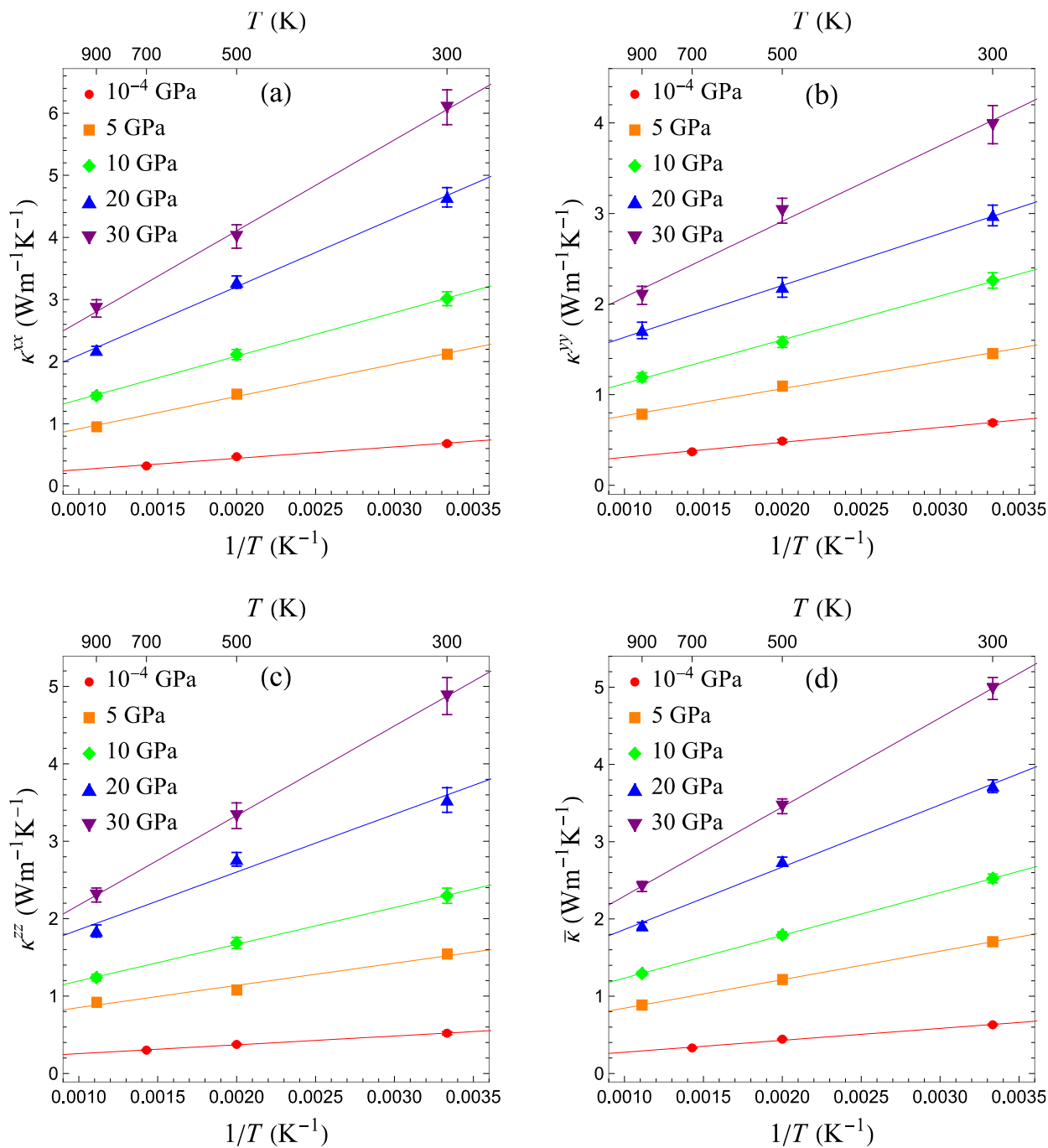

FIG. 3. (a) Thermal conductivity tensor component $\kappa^{xx}$ of β-HMX as a function of reciprocal temperature on the $P = 10^{-4}$, 5, 10, 20, and 30 GPa isobars. Solid lines are linear fits to the data in the $(\kappa, 1/T)$ plane. (b) Same as (a) but for $\kappa^{yy}$. (c) Same as (a) but for $\kappa^{zz}$. (d) Same as (a) but for $\bar{\kappa}$.

## Full Paper

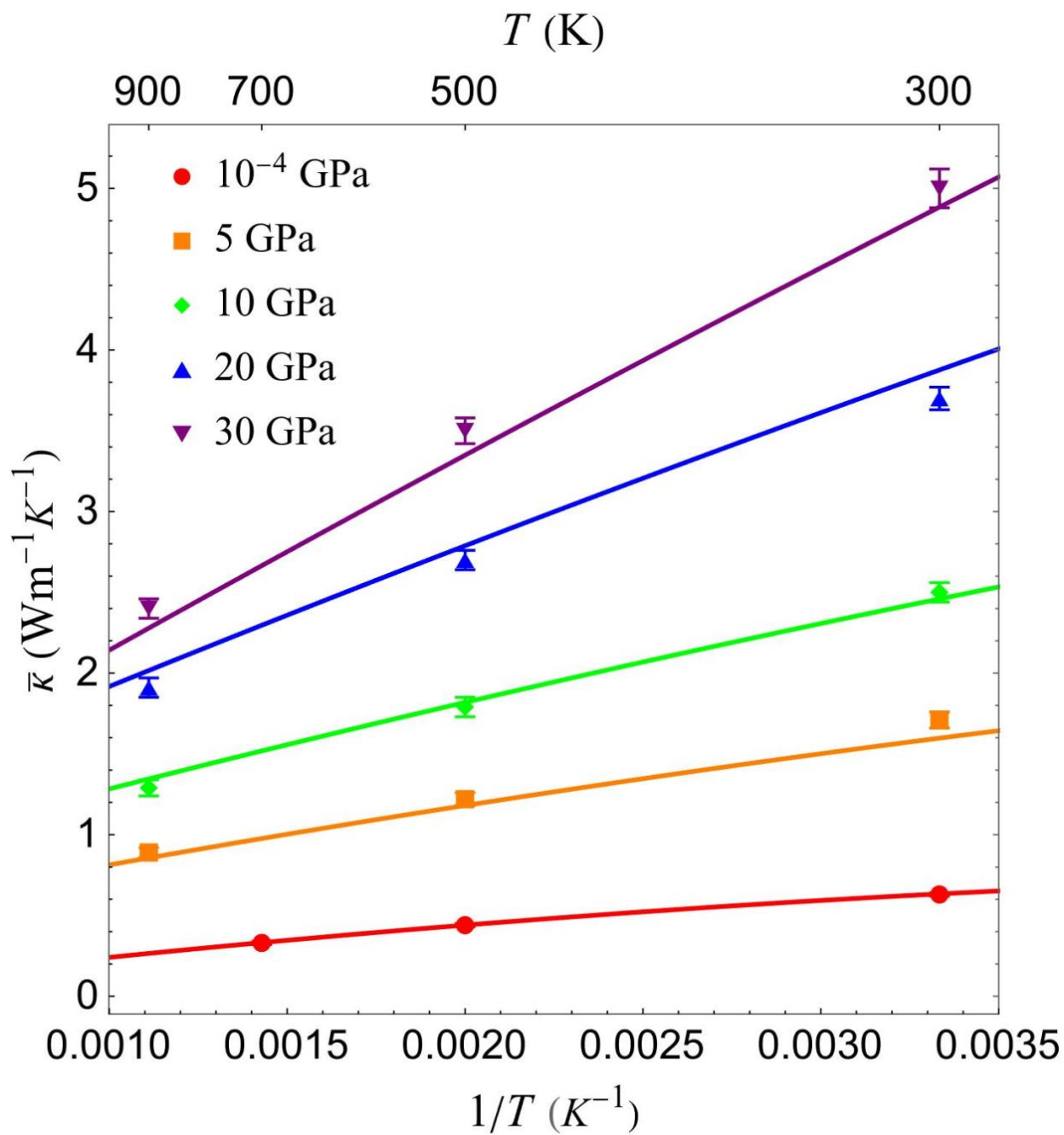

FIG. 4. Predicted $\bar{\kappa} = \bar{\kappa}(T, P)$ values of $\beta$-HMX (symbols) projected onto the ($\bar{\kappa}$, $1/T$) plane for the five isobars indicated in the figure. Solid curves depict the fit of Eq. (7) to the MD predictions.

$-$